\documentclass[letterpaper, 10pt]{article}
\usepackage{preprint}

\title{Ultra-High-Speed Chemiluminescence Tomography of Spinning-Mode Detonation Waves}

\author{
  Amit K. Singh$^1$,
  Mateo Gomez$^2$,
  Kevin Y. Cho$^3$,
  Aaron W. Skiba$^4$,
  and Samuel J. Grauer$^{1,}$\thanks{Corresponding author: \href{mailto:sgrauer@psu.edu}{sgrauer@psu.edu}}\vspace*{.3em}\\
  {\small $^1$Department of Mechanical Engineering, Pennsylvania State University}\vspace*{-.15em}\\
  {\small $^2$Spectral Energies, LLC}\vspace*{-.15em}\\
  {\small $^3$Innovative Scientific Solutions, Inc.}\vspace*{-.15em}\\
  {\small $^4$Aerospace Systems Directorate, Air Force Research Laboratory}\vspace{-1em}}
\date{}

\begin{document}

\maketitle
\setcounter{footnote}{4}
\vspace*{-2em}

\begin{abstract}
\noindent This work presents a chemiluminescence tomography campaign to reconstruct time-resolved, three-dimensional reacting structures in detonation waves propagating through ethylene-based mixtures at 1~atm. Images of chemiluminescence are recorded simultaneously by five cameras through a cylindrical sapphire test section, and a custom calibration procedure is developed to account for refraction through the cylinder. The images are combined to reconstruct an effective emission source term field at megahertz rates. Reconstructions are reported for a persistent spinning detonation, a failed spinning detonation, and a case with counter-propagating transverse fronts. The reconstructed fields enable visualization of wave morphologies, determination of the axial and azimuthal wave speeds, and estimation of key geometric and kinematic parameters. These results demonstrate time-resolved chemiluminescence tomography as a non-intrusive tool for resolving volumetric detonation dynamics that are difficult to infer from point, planar, or line-of-sight diagnostics.\par\vspace{.5em}

\end{abstract}
\vspace*{2em}

\section{Introduction}
\label{sec: introduction}
A striking feature of gaseous detonations is that their mean propagation speed can often be predicted with remarkable accuracy using only conservation laws and equilibrium thermodynamics \cite{Fickett2000, Lee2008}. This success is embodied by the Chapman--Jouguet (CJ) theory, which treats a detonation wave as a discontinuity between the unreacted mixture and equilibrium product state through a one-dimensional (1D) control volume analysis. A more refined picture is provided by the Zel'dovich--von Neumann--D{\"o}ring (ZND) structure, in which a leading shock compresses and heats the mixture and is followed by finite-rate chemical reactions that release heat and thereby support the wave. The ZND model can incorporate detailed kinetics, and extensions of this framework have been used to approximate the effects of heat loss, friction, and curvature \cite{Gelfand1991}. Taken together, the CJ and ZND frameworks provide the foundation for much of detonation theory. Yet detonations observed in the real world are never fully 1D because planar detonations are highly unstable \cite{Lee1984, Shepherd2009}. Instead, detonation waves sustain corrugated fronts comprising Mach stems, incident shocks, and transverse waves, which meet at triple points. Unsteadiness in the shock structure leads to a range of post-shock states wherein heat release may be locally coupled, delayed, or separated from the front \cite{Huang2000, Frederick2025, Zhang2025}. Such multi-dimensional structures play a central role in determining---and therefore in accurately predicting---the propagation limits and stability of detonations in confined geometries.\par

Accurate prediction of detonation behavior is important both for industrial safety \cite{Vasil1999, Gelfand1991} and the design of pressure-gain propulsion concepts, which seek to leverage the thermodynamic advantage associated with near-constant-volume combustion \cite{Paxson2013, Sousa2017, Ma2020, Luan2022}. In both practical and laboratory configurations, detonation propagation shows a marked sensitivity to boundary conditions, including the inflow, confinement geometry, and wall losses, and to chemistry, via the mixture composition and kinetics \cite{Gelfand1991, Shepherd2009, Monnier2022, Crane2023}. As the mixture or geometry approaches a propagation limit, the detonation may no longer sustain a regular multi-headed front and can instead reorganize into ``low-mode propagation,'' including galloping and spinning detonations \cite{Haloua2000}. The spinning mode is a canonical example: a dominant head propagates helically around the wall; its essential structure is organized around a solitary triple point region at which the Mach stem, incident shock, and transverse wave come together \cite{Huang2000, Tsuboi2007, Kitano2009}.\par

Predictive models of such phenomena are needed to understand the propagation limits and stability of detonations, but their development requires experimental measurements that reveal these structures and provide data for validation. Conventional measurements, such as the average wave speed, pressure traces, and soot foil patterns, are invaluable, but they do not fully constrain the instantaneous three-dimensional (3D) arrangement of shock waves and reacting structures. Measurements that recover such 3D structures and can be used to extract key geometric and kinematic factors---for instance, the overall wave shape and the growth or motion of localized reacting features \cite{Fay1952, Jones1976, Huang2000, Tsuboi2007}---and can therefore provide more stringent benchmarks for simulations and reduced-order descriptions.\par

Experimental measurements of detonations are challenging because the relevant dynamics occur over sub-microsecond time scales and involve steep gradients in pressure, temperature, density, and composition. Soot foils, pressure sensors, schlieren and chemiluminescence imaging, and planar fluorescence and scattering measurements have all provided essential insight into detonation structure \cite{Gelfand1991, Pintgen2003, Carter2022, Chavez2024, Frederick2025, Rock2026, Mendiburu2026}. However, these diagnostics are dimensionally limited: they provide wall traces, point histories, planar fields, or line-of-sight integrated projections. This limitation has helped to motivate the use of ``narrow channel'' facilities \cite{Austin2003}, which simplify optical access and the interpretation of image data by restricting out-of-plane motion. Such measurements are valuable for isolating specific features of detonations, particularly those associated with transverse waves and near-limit propagation \cite{Frederick2025, Zhang2025}. At the same time, narrow channels are subject to boundary layer effects, can induce velocity deficits, and preclude the direct observation of propagation modes that require a larger transverse dimension or surface curvature, or that feature a fully 3D topology, including spinning waves propagating in round tubes.\par

Chemiluminescence tomography enables the measurement of these structures \cite{Grauer2023}. In this approach, line-of-sight integrated images recorded from multiple perspectives are combined with an imaging model to reconstruct the underlying 3D emission field. Chemiluminescence measurements are not directly sensitive to state variables (density, temperature, species concentrations), nor do they directly indicate heat release. Nevertheless, they provide a practical marker of reacting structures and can be recorded at the megahertz rates needed to resolve the macroscopic evolution of detonation waves \cite{Athmanathan2022, Frederick2023, Frederick2025}.\par

Tomographic methods have been used extensively in subsonic combustion \cite{Grauer2023}, but their application to detonations remains relatively unexplored. In related work, Gupta et al. \cite{Gupta2024, Gupta2026} used schlieren-based tomography to characterize shock structures in the exhaust of a rotating detonation combustor (RDC), i.e., outside the combustor itself. Gaetano et al.~\cite{Gaetano2021} demonstrated two-view broadband luminescence tomography of reacting structures within an RDC. While they established tomography as a possible route to volumetric imaging of detonations, the two-view setup limited the complexity of the 3D structures that could be recovered. Moreover, curved optical elements, such as cylindrical liners, can warp images of the combustion process, which distorts the reconstructed field unless the imaging model accounts for refraction. Accurate tomographic imaging within an RDC or round detonation tube requires such a model and a reconstruction workflow that is suited to sparse measurements in a cylindrical domain.\par

Here, we demonstrate time-resolved chemiluminescence tomography of detonation waves in an optically accessible round tube. Five MHz-rate cameras image broadband chemiluminescence within a cylindrical test section, and a custom calibration procedure is used to account for refraction through the transparent test section. The emission field is reconstructed using a trilinear basis on a cylindrical grid. We apply the method to three representative cases: a persistent spinning detonation, a spinning detonation that fails, and a detonation with a pair of counter-propagating transverse fronts. This work yields time series of 3D reacting structures that are difficult to isolate with conventional diagnostics and enable the estimation of key geometric and kinematic quantities. In what follows, the fundamentals of chemiluminescence tomography are described in Sec.~\ref{sec: tomography}; the facility and calibration procedure are then presented in Sec.~\ref{sec: facility}; and the measurement campaign is summarized in Sec.~\ref{sec: campaign}. Reconstructions and analysis are presented in Sec.~\ref{sec: results}, followed by our concluding remarks.\par

\section{Chemiluminescence tomography}
\label{sec: tomography}
Chemiluminescence is the spontaneous emission of light from radicals formed in electronically excited states by chemical reactions \cite{Gaydon1974, Eckbreth1987}. Species such as \ce{CH^*}, \ce{OH^*}, \ce{C2^*}, and \ce{CO2^*} are produced by intermediate reactions in combustion and rapidly relax to their ground states, releasing photons in the process. Chemiluminescence generates the blue--green glow observed in many flames and is widely used in experimental studies of combustion because the signal is passive and the resulting images convey information about flame morphology \cite{Geraedts2016, Mohri2017}, stability \cite{Steinberg2012, Yuan2017}, equivalence ratio \cite{Hardalupas2004, Nori2008}, and heat release \cite{Lauer2010, Lauer2011}. However, chemiluminescence images are inherently two-dimensional (2D), loosely corresponding to line-of-sight integrals over the 3D emission source term field within a camera's field of view. When captured from multiple perspectives, these 2D data can be tomographically reconstructed to estimate the underlying source term field. Doing so entails the inversion of a forward imaging model that maps the 3D field to images thereof. This section describes the imaging model, the associated inverse problem, and the optimization method used to solve it.\par

\subsection{Forward model}
\label{sec: tomography: forward}
Imaging of a volumetric emission field can be modeled in terms of a continuum of point sources \cite{Grauer2023}. The signal recorded at the $i$\textsuperscript{th} pixel, $p_i$, is expressed as the convolution of point spread functions throughout a camera's field of view. Let $\boldsymbol{x} \in \mathcal{V}$ denote a 3D point in the probe volume $\mathcal{V}$, and let $\boldsymbol{s} \in \mathcal{S}$ be a 2D point on the sensor plane $\mathcal{S}$. The signal is modeled as
\begin{equation}
    \label{equ: full model}
    p_i = \int_{0}^{\infty}
    \eta_\lambda \,\tau_\lambda
    \int_\mathcal{V}
    g_\lambda(\boldsymbol{x})
    \frac{\Omega\! \left(\boldsymbol{x}\right)}{4\pi}
    B\! \left(\boldsymbol{s}_i, \boldsymbol{x}\right)
    \mathrm{d}\boldsymbol{x} \,\mathrm{d}\lambda,
\end{equation}
where $\eta_\lambda$ is the spectral quantum efficiency and gain of the sensor, $\tau_\lambda$ is the net transmittance of the optics, and $\lambda$ is the wavelength of light. The solid angle subtended by the camera's aperture is $\Omega$, so $\Omega/4\pi$ gives the fraction of light emitted at $\boldsymbol{x}$ that is incident on the sensor. The function $B$ gives the fraction of accepted light that reaches the $i$\textsuperscript{th} pixel, itself centered at $\boldsymbol{s}_i$, as described by Singh et al.~\cite{Singh2025}. The term $g_\lambda$ is the spectral emission source term field.\par

For a single radical species, the source term may be written as
\begin{equation}
    \label{equ: source term}
    g_\lambda(\boldsymbol{x}) =
    \frac{hc}{\lambda}
    \sum_\upsilon \sum_{\ell \in \mathcal{T}_\upsilon}
    \phi_{\upsilon \to \ell}(\lambda)
    \,A_{\upsilon \to \ell}
    \,\widetilde{n}_\upsilon(\boldsymbol{x}),
\end{equation}
where $A_{\upsilon \to \ell}$ is the Einstein $A$ coefficient for spontaneous emission from upper state $\upsilon$ to lower state $\ell$, $\phi_{\upsilon \to \ell}$ is the lineshape function for the same transition, $\mathcal{T}_\upsilon$ is the set of lower states that are accessible from $\upsilon$ through valid transitions, and $\widetilde{n}_\upsilon$ is the local number density of the radical species in state $\upsilon$. The full emission field is obtained by summing Eq.~\eqref{equ: source term} over all emitting species.\par

In principle, Eq.~\eqref{equ: source term} shows that the chemiluminescence spectrum contains information about the population distribution of the emitting radical \cite{Brockhinke2012}. In practice, however, this distribution is difficult to characterize because it depends on the local thermochemical state via competing rates of radical formation, consumption, quenching, and spontaneous emission. Moreover, rotationally resolved spectra from quasi-point probe volumes produce weak signals that are virtually always superimposed on a broadband continuum. Hence, a detailed spectroscopic treatment is impractical for chemiluminescence tomography. Following common practice, we lump detected transitions of the target radical into an effective band-integrated source term, denoted simply as $g$. This field incorporates the spectral weighting by $\eta_\lambda$ and $\tau_\lambda$, the transition probabilities $A_{\upsilon \to \ell}$, and the corresponding upper state populations, such that $g$ is assumed to be proportional to the overall population of the excited radical. However, we do not interpret this field quantitatively, as doing so would require calibration of the optical throughput and detector response, knowledge of the local thermochemical state, and a valid kinetic sub-mechanism for the chemiluminescent species.\par

Assuming that the probe volume is in focus and the pixels are small, the pixel response function $B$ may be approximated by a ray indicator along the 1D path $\mathcal{R}_i \subset \mathcal{V}$ associated with the $i$\textsuperscript{th} pixel \cite{Singh2025}. Invoking the lumped emission model, neglecting blur, and assuming thin rays, Eq.~\eqref{equ: full model} reduces to the effective forward model
\begin{align}
    \label{equ: approximate model}
    p_i = \int_{\mathcal{R}_i} g(\boldsymbol{x}) \,\mathrm{d}\sigma,
\end{align}
where $\mathrm{d}\sigma$ is the differential path length along the $i$\textsuperscript{th} camera ray. Equation~\eqref{equ: approximate model} is used to reconstruct the source term field in arbitrary emission units.\par

\subsection{Tomographic reconstruction}
\label{sec: tomography: reconstruction}
Equation~\eqref{equ: approximate model} describes the signal recorded at a single pixel, where $p_i$ is a ``projection'' of the surrogate emission field $g$. In practice, measurements are available from many pixels across several vantage points, e.g., from distinct cameras or endoscopes. We collect these data into the vector $\boldsymbol{p} = \{p_i\}_{i=1}^{m}$, where $m$ is the total number of pixels of the optical system. The goal of tomographic reconstruction is to estimate a 3D field whose projections are consistent with $\boldsymbol{p}$.\par

\begin{figure}[b]
    \centering
    \includegraphics[width=.6\textwidth]{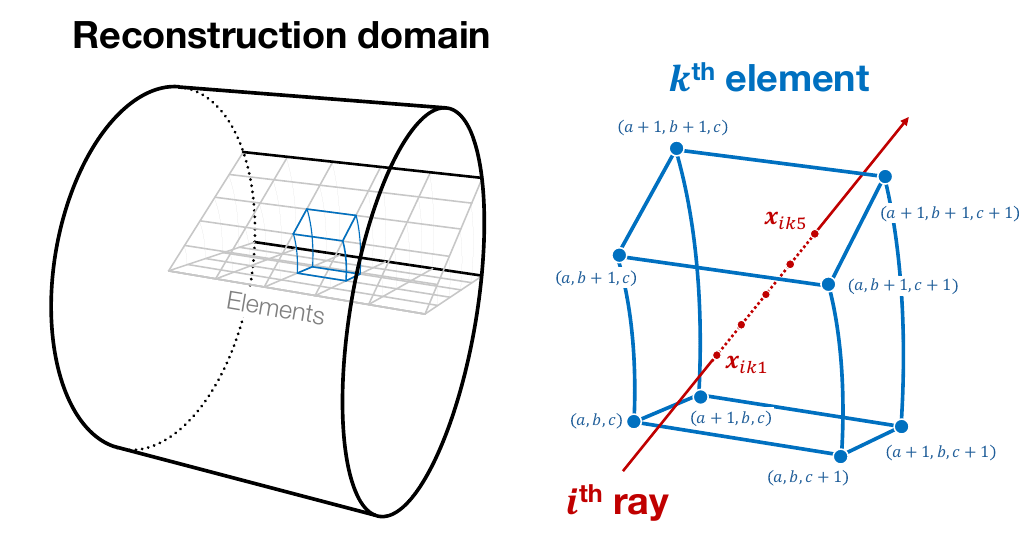}
    \caption{Cylindrical reconstruction grid for the wall-fitted trilinear basis. A representative element is shown to the right.}
    \label{fig: grid}
\end{figure}

Since the unknown field is continuous and thus infinite dimensional, a finite approximation must first be introduced. This is commonly done using voxels, but the detonation structures studied in this work are confined within a cylindrical optical test section and are often attached to the wall. We therefore use a wall-fitted cylindrical basis, which facilitates accurate reconstruction near the boundary. The domain is discretized using radial, azimuthal, and axial nodes, as illustrated in Fig.~\ref{fig: grid}.\par

To define the basis, let $\{r_a\}_{a=1}^{n_r}$, $\{\theta_b\}_{b=1}^{n_\theta}$, and $\{z_c\}_{c=1}^{n_z}$ denote the radial, azimuthal, and axial coordinate nodes, respectively. The azimuthal coordinate is periodic, so $\theta_{n_\theta+1} = \theta_1 + 2\pi$. A ``reconstruction node'' corresponds to a physical $(r, \theta, z)$ location formed from these coordinate nodes and is indexed by a single integer, $j$. For nodes with $r_a > 0$, $j$ corresponds to a unique triplet $(a, b, c)$. The centerline at $r = 0$ is a special case because all the azimuthal locations collapse to the same point. Such nodes are thus represented by a single index $j$ per axial coordinate $(1, -, c)$, with connectivity to all azimuthal sectors in the first radial ring. Hence, the total number of reconstruction nodes is $n = n_z + (n_r - 1) \,n_\theta \,n_z$.\par

The field $g$ is approximated using nodal values $g_j$ and basis functions $\psi_j$ with local support,
\begin{equation}
    \label{equ: discretized 3D field}
    g(\boldsymbol{x}) \approx
    \sum_{j=1}^{n} g_j \,\psi_j(\boldsymbol{x}),
\end{equation}
where $\boldsymbol{g} = \{g_j\}_{j=1}^n$ is the vector of coefficients that specifies the discrete 3D field. Away from the centerline, the basis functions are constructed as tensor products of 1D shape functions,
\begin{equation}
    \label{equ: basis function}
    \psi_{j}(r,\theta,z) =
    \psi_a^{(r)}(r)\;
    \psi_b^{(\theta)}(\theta)\;
    \psi_c^{(z)}(z),
\end{equation}
where $j \leftrightarrow (a,b,c)$, which indicates the correspondence between node $j$ and the radial, azimuthal, and axial indices $a$, $b$, and $c$. The 1D shape functions for $\alpha \in \{r, \theta, z\}$ at interior nodes are given by
\begin{equation}
    \label{equ: linear shape function}
    \psi_q^{(\alpha)}(\alpha) =
    \begin{cases}
    \frac{\alpha-\alpha_{q-1}}{\alpha_q-\alpha_{q-1}}, & \alpha_{q-1}<\alpha\le\alpha_q, \\
    \frac{\alpha_{q+1}-\alpha}{\alpha_{q+1}-\alpha_q}, & \alpha_q<\alpha\le\alpha_{q+1}, \\
    0, & \text{otherwise},
\end{cases}
\end{equation}
with $q \in \{1, \dots, n_\alpha\}$. For non-periodic directions $r$ and $z$, the end nodes $q = 1$ and $q = n_\alpha$ use the one-sided form obtained by omitting the branch outside the domain. In the azimuthal direction, the same expression is applied with periodic indexing. At the centerline, the radial interpolation is one-sided and the nodal value is shared across all azimuthal sectors because $\theta$ is degenerate at $r = 0$.\par

Substituting Eq.~\eqref{equ: discretized 3D field} into Eq.~\eqref{equ: approximate model} gives
\begin{equation}
    \label{equ: ray sum equation}
    p_i \approx
    \int_{\mathcal{R}_i}
    \sum_{j=1}^{n} g_j \,\psi_j\! \left[\boldsymbol{x}(\sigma)\right]
    \mathrm{d}\sigma =
    \sum_{j=1}^{n} A_{ij} \,g_j,
\end{equation}
where $A_{ij}$ is the sensitivity of the $i$\textsuperscript{th} ray to the $j$\textsuperscript{th} reconstruction node,
\begin{equation}
    \label{equ: node sensitivity}
    A_{ij} =
    \frac{\partial p_i}{\partial g_j} =
    \int_{\mathcal{R}_i}
    \psi_j\! \left[\boldsymbol{x}(\sigma)\right]
    \mathrm{d}\sigma.
\end{equation}
This is the standard algebraic formulation commonly used in tomography \cite{Gordon1970, Verhoeven1993, Grauer2023}. The present implementation differs from prior work on combustion tomography primarily through its use of trilinear interpolation on a wall-fitted cylindrical grid. As shown in Fig.~\ref{fig: grid}, this grid is composed of 3D elements, indexed by $k$, that are bounded by adjacent coordinate nodes. Let $\mathcal{K}_j$ denote the set of elements adjacent to node $j$. Since $\psi_j = 0$ outside the union of these elements, Eq.~\eqref{equ: node sensitivity} need only be evaluated over $\mathcal{K}_j$. Interior off-axis nodes, with $0 < r_a < r_{n_r}$ and $z_1 < z_c < z_{n_z}$, are adjacent to eight elements, while nodes on the outer radial boundary, axial boundaries, or centerline have modified connectivity.\par

The sensitivity coefficients are computed by ray tracing. For the $i$\textsuperscript{th} ray and $k$\textsuperscript{th} element, the ray--element intersection defines a segment of length $\Delta \sigma_{ik}$, with $\Delta \sigma_{ik} = 0$ if the ray does not intersect the element. The integral over each non-zero segment, contained in the subset $\mathcal{K}_{ij} \subseteq \mathcal{K}_j$, is evaluated using five-point Gauss--Legendre quadrature:
\begin{equation}
    \label{equ: GLQ integral}
    A_{ij} \approx
    \sum_{k \in \mathcal{K}_{ij}}
    \frac{\Delta\sigma_{ik}}{2}
    \sum_{q=1}^{5}
    w_q \,\psi_j(\boldsymbol{x}_{ikq}),
\end{equation}
where $w_q$ are the weights and $\boldsymbol{x}_{ikq}$ are the quadrature points along the intersection of ray $i$ with element $k$, as shown in Fig.~\ref{fig: grid}. The factor of $\Delta \sigma_{ik}/2$ accounts for the mapping from the standard quadrature interval $[-1,1]$ to the physical ray segment. Repeating this procedure for all rays and reconstruction nodes yields the sensitivity matrix $\boldsymbol{A} \in \mathbb{R}^{m\times n}$ and the discrete forward model
\begin{equation}
    \label{equ: linear equation}
    \boldsymbol{p} \approx \boldsymbol{Ag}.
\end{equation}
The corresponding ``data loss'' is
\begin{equation}
    \label{equ: data loss}
    \mathscr{J}_\mathrm{data} =
    \frac{1}{2}
    \left\lVert\boldsymbol{Ag} - \boldsymbol{p}\right\rVert_2^2,
\end{equation}
which is minimized when projections of the field given by $\boldsymbol{g}$ are consistent with the image data in $\boldsymbol{p}$.\par

For a high-resolution basis, the number of unknown coefficients is large relative to the number of independent projections. This ``limited-data'' tomography problem is inherently underdetermined. Although the basis could be coarsened to reduce the number of unknowns, doing so would introduce significant model error and compromise the reconstruction of wall-attached structures. Instead, we regularize the inverse problem using a second-order Tikhonov penalty,
\begin{equation}
    \label{equ: continuous regularization}
    \mathscr{J}_\mathrm{reg} =
    \frac{1}{2} \int_\mathcal{V}
    \left\lVert \nabla_{\boldsymbol{x}}^2 g(\boldsymbol{x}) \right\rVert_2^2
    \mathrm{d}\boldsymbol{x} \approx
    \frac{1}{2} \left\lVert \boldsymbol{L}\boldsymbol{g} \right\rVert_2^2,
\end{equation}
where $\nabla_{\boldsymbol{x}}^2$ is the spatial Laplacian operator and $\boldsymbol{L}$ is its discrete approximation, implemented using second-order spatial derivatives in each coordinate direction. This penalty favors smooth fields with low curvature and suppresses artifacts in the null space of $\boldsymbol{A}$.\par

The reconstruction is obtained by solving
\begin{equation}
    \label{equ: aggregate loss}
    \boldsymbol{g}^{\star} =
    \arg\min_{\boldsymbol{g} \ge 0}
    \left(
        \mathscr{J}_\mathrm{data} +
        \gamma^2 \mathscr{J}_\mathrm{reg}
    \right) \!.
\end{equation}
The weighting parameter $\gamma$ controls the trade-off between fitting the measured projections and enforcing smoothness. When $\gamma$ is small, the objective is dominated by the data term and can permit null-space artifacts. Conversely, when $\gamma$ is large, the solution is dominated by the regularization term and becomes overly smooth. We select $\gamma$ using an L-curve analysis \cite{Hansen2001}. The constraint $\boldsymbol{g} \ge 0$ enforces non-negativity of the source term field.\par

\subsection{Optimization method}
\label{sec: tomography: optimization}
The reconstruction problem defined in Eq.~\eqref{equ: aggregate loss} is a large-scale, non-negative least-squares problem. The objective is quadratic in $\boldsymbol{g}$, and its gradient is
\begin{equation}
    \label{equ: objective gradient}
    \boldsymbol{h}(\boldsymbol{g}) =
    \nabla_{\boldsymbol{g}} \mathscr{J}(\boldsymbol{g}) =
    \boldsymbol{A}^\top
    \left(\boldsymbol{Ag} - \boldsymbol{p}\right) +
    \gamma^2 \boldsymbol{L}^\top \boldsymbol{Lg}.
\end{equation}
Direct formation and inversion of the Hessian are impractical for the problem sizes considered here, so we minimize Eq.~\eqref{equ: aggregate loss} using a limited-memory Broyden--Fletcher--Goldfarb--Shanno (L-BFGS) algorithm \cite{Nocedal2006}.\par

At iteration $k$, the current estimate $\boldsymbol{g}_k$ is updated according to
\begin{equation}
    \label{equ: projected update}
    \boldsymbol{g}_{k+1} =
    \mathsf{P}_+
    \left(
        \boldsymbol{g}_{k} +
        t_k \boldsymbol{d}_k
    \right) \!,
\end{equation}
where $\boldsymbol{d}_k$ is the search direction, $t_k$ is the step size, and the operator $\mathsf{P}_+$ projects the argument onto the non-negative orthant, defined element-wise as
\begin{equation}
    \label{equ: positivity projection}
    \left[\mathsf{P}_+(\boldsymbol{g})\right]_j =
    \max(g_j, 0).
\end{equation}
The search direction is scaled using an approximation to the inverse Hessian,
\begin{equation}
    \label{equ: BFGS direction}
    \boldsymbol{d}_k =
    -\boldsymbol{B}_k \widetilde{\boldsymbol{h}}_k,
\end{equation}
where $\boldsymbol{B}_k \approx [\nabla_{\boldsymbol{g}}^2 \mathscr{J}(\boldsymbol{g}_k)]^{-1}$ and $\widetilde{\boldsymbol{h}}_k = \{h_{kj}\}_{j=1}^n$ is the projected gradient that accounts for the non-negativity constraint.\par

To compute $\widetilde{\boldsymbol{h}}_k$, an ``active set'' identifies nodes for which $g_j$ is already at the lower bound and an unconstrained step would decrease it below zero:
\begin{equation}
    \label{equ: active set}
    \mathcal{A}_k =
    \{j \mid g_{kj} = 0, \;h_{kj} > 0\},
\end{equation}
where $h_{kj}$ is the $j$\textsuperscript{th} component of $\boldsymbol{h}(\boldsymbol{g}_k)$ from Eq.~\eqref{equ: objective gradient}. The projected gradient is then
\begin{equation}
    \label{equ: projected gradient}
    \widetilde{h}_{kj} =
    \begin{cases}
        0, & j \in \mathcal{A}_k, \\
        h_{kj}, & \text{otherwise}.
    \end{cases}
\end{equation}
Equation~\eqref{equ: projected gradient} freezes active nodes in the BFGS search direction while allowing free nodes to evolve according to the standard update.\par

The inverse Hessian approximation is implicitly defined using the most recent curvature pairs,
\begin{subequations}
    \label{equ: curvature information}
    \begin{align}
        \boldsymbol{u}_k &= \boldsymbol{g}_{k+1} - \boldsymbol{g}_k, \\
        \boldsymbol{y}_k &= \boldsymbol{h}(\boldsymbol{g}_{k+1}) - \boldsymbol{h}(\boldsymbol{g}_k),
    \end{align}
\end{subequations}
where $\boldsymbol{u}_k$ is the update produced by Eq.~\eqref{equ: projected update} and $\boldsymbol{y}_k$ is the change in the gradient. A pair is retained only if it satisfies the curvature condition, $\boldsymbol{u}_k^\top \boldsymbol{y}_k > 0$; otherwise, it is discarded. In full-memory BFGS, these pairs define the inverse Hessian update as
\begin{equation}
    \label{equ: inverse hessian update}
    \boldsymbol{B}_{k+1} =
    \left(\boldsymbol{I} - \rho_k \boldsymbol{u}_k \boldsymbol{y}_k^\top\right)
    \boldsymbol{B}_{k}
    \left(\boldsymbol{I} - \rho_k \boldsymbol{y}_k \boldsymbol{u}_k^\top\right) +
    \rho_k \boldsymbol{u}_k \boldsymbol{u}_k^\top,
\end{equation}
where $\rho_k = (\boldsymbol{y}_k^\top \boldsymbol{u}_k)^{-1}$. In L-BFGS, however, $\boldsymbol{B}_k$ is not stored explicitly. Rather, the product $\boldsymbol{B}_k \widetilde{\boldsymbol{h}}_k$ in Eq.~\eqref{equ: BFGS direction} is evaluated by applying the standard two-loop recursion over the retained $(\boldsymbol{u}_k, \boldsymbol{y}_k)$ pairs.\par

The step size $t_k$ is selected using a line search with decrease and curvature conditions applied along the projected path in Eq.~\eqref{equ: projected update}. We use line-search constants of $c_1 = 10^{-4}$ and $c_2 = 0.4$ and a history of 30 $(\boldsymbol{u}_k, \boldsymbol{y}_k)$ pairs to compute $\boldsymbol{B}_k \widetilde{\boldsymbol{h}}_k$.\par

\section{Facility and camera calibration}
\label{sec: facility}
\begin{figure*}[ht]
    \centering
    \includegraphics[width=\textwidth]{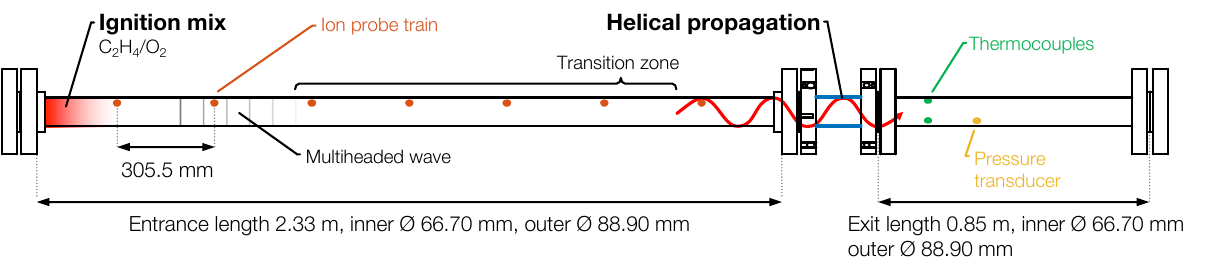}
    \caption{Schematic of the optically accessible detonation tube and diagnostics.}
    \label{fig: detonation tube}
\end{figure*}

Experiments were performed in an optically accessible round tube in the Detonation Engine Research Facility at AFRL \cite{Gomez2024}, shown schematically in Fig.~\ref{fig: detonation tube}. We next describe the tube, imaging arrangement, and camera calibration procedure used to account for refraction through the cylindrical optical test-section.\par

\subsection{Optically accessible round tube}
\label{sec: facility: detonation tube}
The detonation tube consists of a pre-detonator, a driven section, an optically accessible test section, and an extended downstream section. Spark plugs in the pre-detonator fire into a stoichiometric \ce{C2H4}--\ce{O2} ``ignition mixture'' to initiate a self-sustained detonation; a Shchelkin spiral promotes repeatable initiation and rapid deflagration-to-detonation transition. The 2.33~m driven section contains ion probes spaced at 305.5~mm intervals to measure the detonation wave speed, although only the first and last probe pairs were actively used for the present experiments. Optical access is provided by a cylindrical sapphire test section with ID~66.70~mm, OD~88.90~mm, and length~139.7~mm. The extended downstream section reduces end effects and houses thermocouples, pressure sensors, and soot foils for auxiliary monitoring.\par

\begin{figure}[ht]
    \centering
    \includegraphics[width=.475\textwidth]{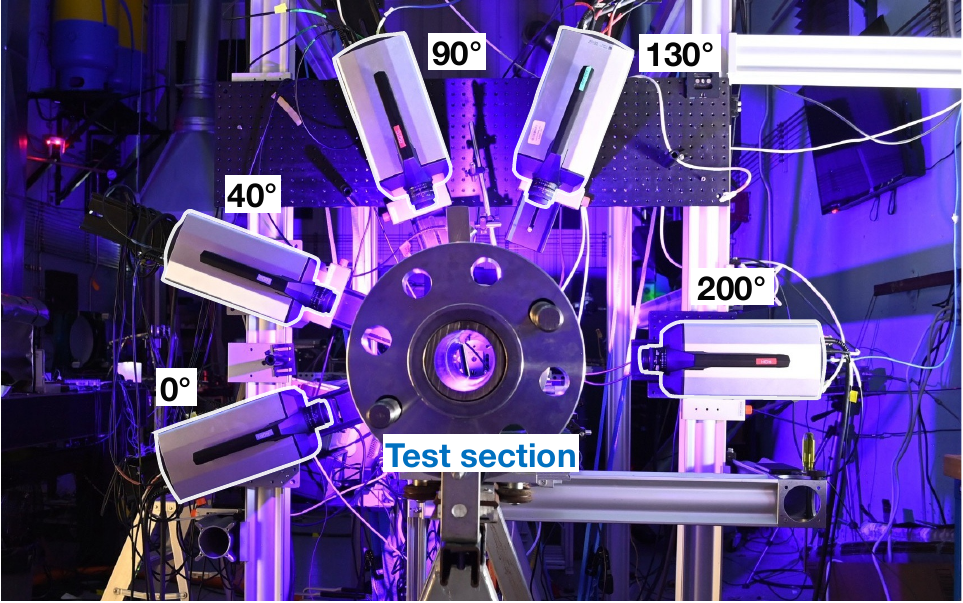}
    \caption{Top view of camera placement around the test section.}
    \label{fig: camera setup}
\end{figure}

Upon initiation, the detonation wave propagates along the tube and passes through the visible test section. Chemiluminescence is recorded during this passage using five high-speed cameras positioned around the tube at a distance of roughly 380~mm from the center and at angles of 0\textdegree, 40\textdegree, 90\textdegree, 130\textdegree, and 200\textdegree, as shown in Fig.~\ref{fig: camera setup}. Each camera is equipped with a 50~mm lens. The cameras are triggered by a photodiode placed just upstream of the test section, and they are synchronized to within 10~ns. Images are acquired at frame rates ranging from 2--5~MHz, using an exposure of 110--200~ns to freeze the wave and aperture settings between $f/2.8$ and $f/11$; 128 frames are recorded for each detonation. No optical filters are used; however, \ce{OH^*} emission is largely rejected by the lenses because they do not transmit UV light. The detected signal is therefore dominated by \ce{CH^*}, with lesser contributions from \ce{C2^*} for non-rich cases as well as broadband light from triatomic radicals.\par

\subsection{Camera calibration procedure}
\label{sec: facility: calibration}
Tomographic reconstruction requires an accurate camera model because the ray paths $\mathcal{R}_i$ determine the structure of $\boldsymbol{A}$ and thus the fidelity of Eq.~\eqref{equ: linear equation}. Standard methods based on planar targets, such as Zhang's method \cite{Zhang2002}, model each camera as a perspective system subject to radial and tangential distortion by the lens, but they do not account for refraction through curved optical elements. This treatment is insufficient for the present application because images of the detonation tube are warped by refraction through the optical section. We therefore adapt the approach of Paolillo and Astarita \cite{Paolillo2020, Liu2019}, using a custom multi-camera calibration framework that enforces Snell's law as rays pass through the outer and inner surfaces of the cylinder.\par

\begin{figure}[ht]
    \centering
    \includegraphics[width=.65\textwidth]{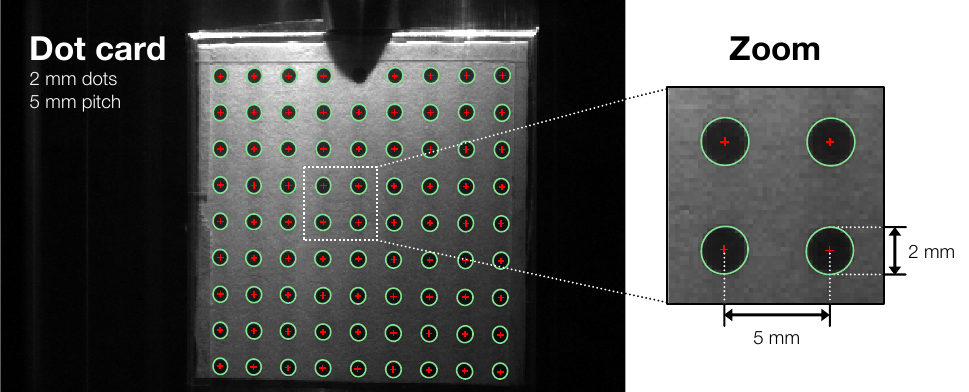}
    \caption{Sample image of an illuminated calibration target with the cylinder in place, along with a zoomed view. Detected points are shown in green and reprojected points are shown in red.}
    \label{fig: calibration target}
\end{figure}

The calibration data are obtained using a custom planar target consisting of a $9\times 9$ grid of black dots, 2~mm in diameter and separated by a 5~mm pitch, printed on a white background. Relative dot locations on the target are known, so they provide control points against which the imaging model can be evaluated. The target is mounted on a multi-axis stage and imaged at a range of translations and orientations. For each pose, images of the dot card are recorded by all the cameras, with the target face visible to a subset thereof. Dot centroids are extracted using an in-house MATLAB algorithm, the results of which are illustrated by the green markers in Fig.~\ref{fig: calibration target}. To limit the effects of chromatic aberration on the calibration, the target is illuminated using a narrow-band LED centered at 405~nm, which is near the detected \ce{CH^*} band.\par

\begin{figure}[b]
    \centering
    \includegraphics[width=.475\textwidth]{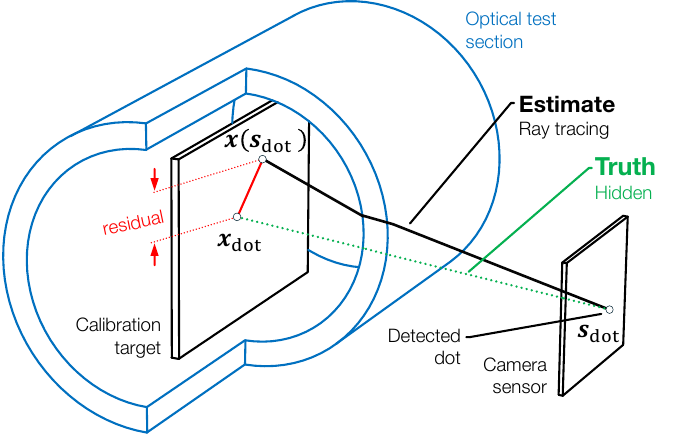}
    \caption{Illustration of the calibration framework.}
    \label{fig: calibration}
\end{figure}

The calibration proceeds in two stages. In the first stage, images recorded without the cylinder are used to estimate the camera intrinsics, radial and tangential distortion coefficients, and relative camera poses. Zhang's method is applied to each camera \cite{Zhang2002}, and simultaneous images of the dot card from different views are used to place the camera models into a shared coordinate system. In the second stage, images of the dot card recorded through the cylinder are used to calibrate the refractive imaging model. The intrinsics, distortion coefficients, and relative camera poses from the first stage are retained. The remaining unknowns are the pose of the full camera assembly relative to the cylinder, the pose of the dot card in each image, and the cylinder parameters, including its inner radius, outer radius, and refractive index. The global coordinate system is defined with respect to the base of the cylinder and its axis.\par

Figure~\ref{fig: calibration} illustrates the calibration procedure. For a dot detected at the sensor location $\boldsymbol{s}_\mathrm{dot}$, the corresponding principal ray is traced from the camera through the cylinder, applying Snell's law at each interface. The ray is then intersected with the estimated target plane to predict the physical dot location, $\boldsymbol{x}(\boldsymbol{s}_\mathrm{dot})$, which is compared to the dot location implied by the estimated camera position and target pose, $\boldsymbol{x}_\mathrm{dot}$. The residuals $\|\boldsymbol{x}_\mathrm{dot} - \boldsymbol{x}(\boldsymbol{s}_\mathrm{dot})\|_2^2$ are summed over all the visible dots for all cameras and calibration images. The unknown calibration parameters are determined by minimizing this objective using the Levenberg--Marquardt algorithm. For stability, the optimization is performed sequentially: first, the camera assembly and target poses are optimized while the cylinder radii and refractive index are fixed at their nominal values; then, all the parameters are optimized together. The red markers in Fig.~\ref{fig: calibration target} show the reprojected dot locations obtained from the final optimized model.\par

Calibration results are summarized in Table~\ref{tab: calibration}, which lists the calibrated focal lengths $f$, distances from each camera to the cylinder axis $d$, and root-mean-square reprojection errors $\varepsilon$. The reprojection errors are approximately 0.15--0.3~px, indicating good agreement between the detected and reprojected dot locations. The largest error occurs for the camera positioned at 200\textdegree\, which we attribute to the limited number of well-illuminated dot card images available for that view. Errors were also somewhat larger near the edges of the card, where non-uniform illumination degraded the accuracy of dot detection. The calibrated focal lengths are within 2\% of the nominal 50~mm lens specification. The camera distances are close to the expected value of 380~mm, with the variation in $d$ attributed primarily to genuine differences in the camera-to-axis distances. The optimized inner and outer cylinder radii, 33.33~mm and 44.49~mm, agree with direct measurements to within 0.02~mm, and the calibrated refractive index of 1.76 is reasonably consistent with the expected value for sapphire at 405~nm, i.e., 1.81. Taken together, these results support the accuracy of the ray paths used in our tomographic reconstructions.\par

\begin{table}[ht]
    \footnotesize
    \centering
    \caption{Overview of camera calibration results.}
    \vspace*{.5em}
    \begin{tabular}{c c c c}
        \hline
        Camera position & $f$,~mm & $d$,~mm & $\varepsilon$,~px \\
        \hline
        0\textdegree\   & 50.87 & 382.45 & 0.179 \\
        40\textdegree\  & 50.81 & 390.07 & 0.167 \\
        90\textdegree\  & 50.54 & 379.73 & 0.223 \\
        130\textdegree\ & 50.68 & 382.74 & 0.179 \\ 
        200\textdegree\ & 50.86 & 372.44 & 0.300 \\ 
        \hline
    \end{tabular}
    \label{tab: calibration}
\end{table}

\section{Experimental cases and wave structures}
\label{sec: campaign}
The experimental campaign generated a range of detonation morphologies, including single-, dual-, and multi-headed waves. We focus on three cases that show how chemiluminescence tomography can resolve wave structures that are difficult to ascertain from 2D projection data. The campaign and selected cases are summarized below, followed by a brief discussion of the geometric and kinematic parameters used to interpret the single-head waves.\par

\subsection{Campaign overview}
\label{sec: campaign: overview}
The full campaign comprised 47 detonations spanning three mixture families: \ce{C2H4}--air--\ce{Ar}, \ce{C2H4}--air, and \ce{C2H4}--\ce{O2}--\ce{Ar}. Here, ``air'' refers to a 0.79\slash 0.21 mixture of \ce{N2}--\ce{O2} by mole fraction. The \ce{C2H4}--air--\ce{Ar} cases were run near stoichiometric conditions with \ce{Ar} mole fractions from approximately 0.35 to 0.65; the \ce{C2H4}--air cases spanned lean to highly fuel-rich conditions, with $0.65 \leq \widetilde{\phi} \leq 3.5$. Initial pressures were close to atmospheric, and initial temperatures were near ambient. Measured wave speeds were generally close to the corresponding CJ speeds, $D_\mathrm{CJ}$, and the observed morphologies included non-spinning fronts, persistent and quasi-spinning waves, multi-headed waves, and wave failure within the test section.\par

\begin{figure*}[t]
    \centering
    \includegraphics[width=\textwidth]{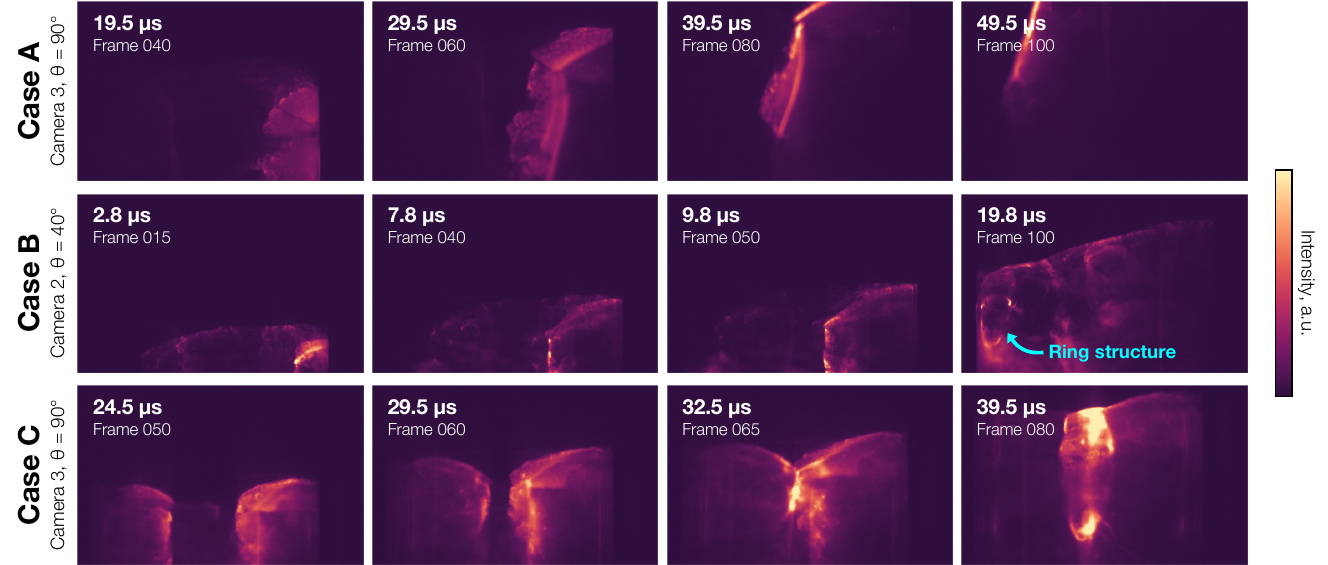}
    \caption{Projections for the selected cases: (a)~persistent spinning detonation in a near-stoichiometric \ce{C2H4}--air--\ce{Ar} mixture, (b)~spinning detonation followed by wave failure in a fuel-rich \ce{C2H4}--air mixture, and (c)~counter-propagating transverse detonations in a lean \ce{C2H4}--air mixture.}
    \label{fig: projections}
\end{figure*}

\begin{table*}[!ht]
    \footnotesize
    \centering
    \caption{Detonation parameters: equivalence ratio $\widetilde{\phi}$, \ce{Ar} mole fraction $X_{\ce{Ar}}$, ambient $(\cdot)_0$ and von Neumann $(\cdot)_\mathrm{VN}$ state variables, CJ speed $D_\mathrm{CJ}$, induction time $\tau_\mathrm{ind}$, and exothermic time $\tau_\mathrm{ex}$.}
    \vspace*{.5em}
    \begin{tabular}{c c c c c c c c c c c}
        \hline \noalign{\vskip 2pt}
        Case &
        Mixture &
        $\widetilde{\phi}$ &
        $X_{\ce{Ar}}$ &
        $p_0$,~kPa &
        $T_0$,~K &
        $p_\mathrm{VN}$,~MPa &
        $T_\mathrm{VN}$,~K &
        $D_\mathrm{CJ}$,~m/s &
        $\tau_\mathrm{ind}$,~$\upmu$s &
        $\tau_\mathrm{ex}$,~$\upmu$s \\
        \hline \noalign{\vskip 2pt}
        A & \ce{C2H4}--air--\ce{Ar} & 1.0 & 0.65 & 99.1 & 307 & 2.26 & 1618 & 1418 & 5.4 & 1.5 \\
        B & \ce{C2H4}--\ce{O2}--\ce{Ar} & 3.5 & 0.63 & 99.2 & 305 & 3.35 & 1634 & 1618 & 1.0 & 1.2 \\
        C & \ce{C2H4}--air & 0.65 & 0 & 99.1 & 309 & 2.66 & 1442 & 1671 & 6.1 & 0.6 \\
        \hline
    \end{tabular}
    \label{tab: cases}
\end{table*}

\subsection{Selected cases}
\label{sec: campaign: cases}
Three cases are selected for tomographic analysis, with key parameters summarized in Table~\ref{tab: cases}. The detonation parameters were computed using the Shock and Detonation Toolbox \cite{Kao2023} with the FFCM-2 mechanism \cite{Zhang2023}. We use the measured wave speeds relative to the CJ speed to gauge the velocity deficit; the induction and exothermic times, $\tau_\mathrm{ind}$ and $\tau_\mathrm{ex}$, indicate how tightly heat release is coupled to shock compression.\par

Case~A is a persistent spinning detonation in a near-stoichiometric \ce{C2H4}--air--\ce{Ar} mixture. The wave propagates slightly below the computed CJ speed, consistent with the modest velocity deficits observed in many near-limit detonations. Projection data from Case~A, presented in the top row of Fig.~\ref{fig: projections}, show a coherent luminous structure propagating from right to left through the test section. The wave contains a sharp wall-attached front and an oblique structure that penetrates radially inward toward the tube axis. This morphology is recognizably that of a single-head spinning detonation, but the full 3D arrangement of the Mach stem, transverse wave, and their inward extent is ambiguous via any given image.\par

Case~B is a highly fuel-rich spinning detonation that fails within the test section. Early in the sequence, shown in the middle row of Fig.~\ref{fig: projections}, the wave has a spinning structure that is broadly similar to that in Case~A, although the luminous front is finer and less coherent. Between the frames recorded at 9.8~$\upmu$s and 19.8~$\upmu$s, the detonation fragments and the spinning mode fails. Ring-like structures appear to emerge from the wall, superimposed on distributed emission that extends across much of the tube. These features, discussed in the results section, are difficult to isolate in projection images due to the line-of-sight integrated nature of chemiluminescence data.\par

Case~C is a lean \ce{C2H4}--air detonation that features two counter-propagating transverse detonations. The measured speed is below the computed CJ speed, and the relatively long induction time suggests a greater separation between shock compression and heat release. As shown in the bottom row of Fig.~\ref{fig: projections}, two intense transverse fronts approach one another, collide, and then fragment into (relatively) fast- and slow-moving structures that propagate axially.\par

\subsection{Spinning-mode detonation wave parameters}
\label{sec: campaign: spinning}
\begin{figure}[t]
    \centering
    \includegraphics[width=.4\textwidth]{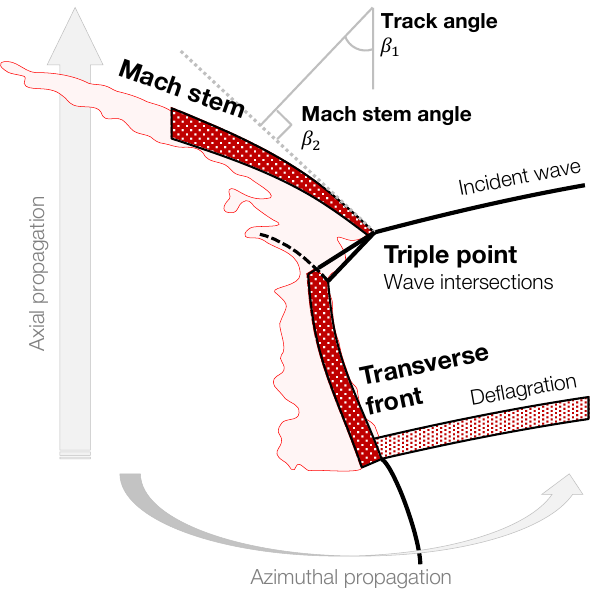}
    \caption{Idealized structure of a wall-attached, spinning-mode detonation wave, adapted from Huang et al.~\cite{Huang2000}. Red zones mark regions with combustion reactions.}
    \label{fig: spinning structure}
\end{figure}

Case~A is interpreted using the standard geometric description of single-headed spinning detonation, developed at length in prior experimental and numerical studies \cite{Huang2000, Kitano2009, Zhang2024}. In this description, a single triple point propagates helically around the tube. The local wave structure consists of a Mach stem, an incident front, and a transverse wave, which converge at the triple point, as illustrated in Fig.~\ref{fig: spinning structure}. Heat release is concentrated behind the Mach stem and transverse wave. The weaker incident front leaves a shocked mixture in its wake, wherein heat release is delayed or spatially separated from the front; this shocked mixture helps to support the azimuthal propagation of the transverse detonation by reducing $\tau_\mathrm{ind}$ \cite{Frederick2025}. Because we measure chemiluminescence rather than density or pressure, the incident front is not observed. We must therefore educe the spinning structure from the heat release zones associated with the Mach stem and transverse wave and their helical motion.\par

Several quantities are commonly used to describe spinning detonations. For instance, the pitch $P$ is the axial distance traveled by the spinning head during a full circumferential rotation, and the track angle $\beta_1$ is the angle between the resulting helical trajectory and the axial direction. These quantities are related by
\begin{equation}
    \label{equ: spinning track angle}
    \cot(\beta_1) =
    \frac{D_\mathrm{ax}}{D_\theta} =
    \frac{P}{\pi D},
\end{equation}
where $D_\mathrm{ax}$ and $D_\theta$ are the axial and azimuthal components of the wave velocity and $D$ is the tube diameter. A second angle, $\beta_2$, characterizes the local orientation of the Mach stem. Specifically, $\beta_2$ is the angle between the incoming flow and the Mach stem in a shock-attached frame of reference. Here, $\beta_1$ and $\beta_2$ correspond to $\alpha$ and $\Phi_3$, respectively, in the notation of Huang et al.~\cite{Huang2000}. Values of $\beta_2$ near 90\textdegree\ indicate that the Mach stem is nearly normal to the incoming flow and therefore represents the strongest portion of the leading front. Prior work reports pitch values of order $3D$, track angles $\beta_1$ of 42--49\textdegree, and Mach-stem angles $\beta_2$ close to 90\textdegree\ for persistent spinning waves \cite{Fay1952, Huang2000, Tsuboi2007, Kitano2009}. In the results section, we use the reconstructed chemiluminescence fields to identify the 3D wave trajectory and estimate $\beta_1$ and $\beta_2$.\par

\section{Results and discussion}
\label{sec: results}
We now present reconstructions for the selected cases. To start, we estimate key wave parameters for the persistent spinning detonation. After that, we consider the other cases, which feature detonation failure and counter-propagating transverse fronts.\par

The common volume observed by all cameras is determined from the calibrated model. It is approximately 67~mm in diameter and 50~mm in height. This region is discretized using $50 \times 520 \times 100$ points in the radial, azimuthal, and axial directions, giving a characteristic spacing below 1~mm for each direction. As such, the discrete imaging model remains accurate whilst keeping memory usage in check. The sensitivity matrix $\boldsymbol{A}$ and discrete Laplacian $\boldsymbol{L}$ are precomputed. Pixels that do not image the reconstruction domain are removed from the problem, and the resulting system is solved using the L-BFGS procedure described in Sec.~\ref{sec: tomography: optimization}. L-curve analysis is performed for each dataset, and a regularization weight of $\gamma = 0.45$ is used for all the reconstructions presented here.\par

\subsection{Case~A: persistent spinning detonation}
\label{sec: results: persistent-spin}
\begin{figure*}[t]
    \centering
    \includegraphics[width=.9\textwidth]{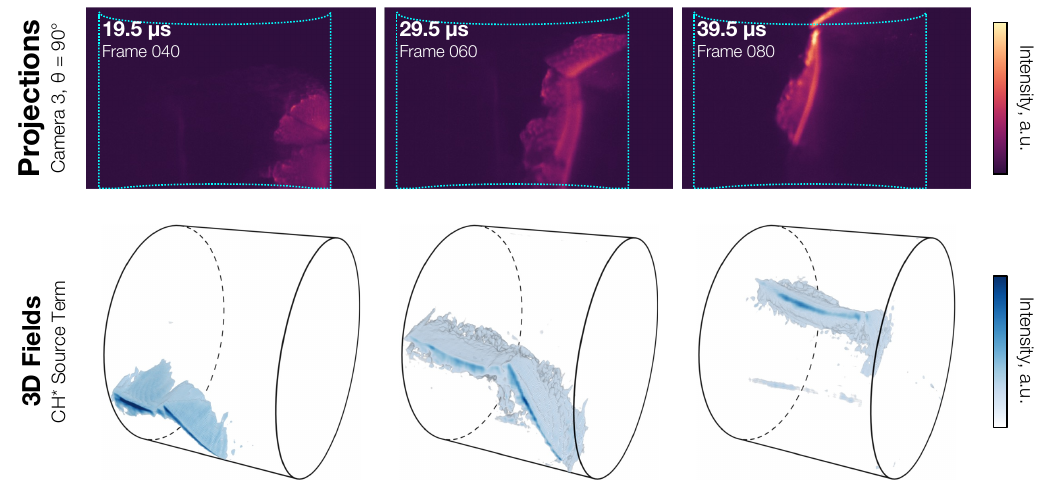}
    \caption{Projections (top) and reconstructions (bottom) of a spinning detonation in a near-stoichiometric \ce{C2H4}--air--\ce{Ar} mixture. Dashed teal lines superimposed on the projections indicate the boundary of the reconstruction volume.}
    \label{fig: reconstructions: case A}
\end{figure*}

Figure~\ref{fig: reconstructions: case A} presents reconstructed chemiluminescence fields for Case~A at $t = 19.5~\upmu$s, 29.5~$\upmu$s, and 39.5~$\upmu$s (bottom row), together with the corresponding projection images from one of the cameras (top row). The first two frames show the spinning head propagating toward the camera and leftwards; the final frame shows a side view of the wave. The reconstruction is consistent with the standard interpretation of a spinning detonation. A strong wall-attached front is visible---extending radially inward from the wall to about $0.5R$, consistent with prior assessments of single-head spinning detonations \cite{Tsuboi2007}---together with an oblique reacting structure that stretches downward. The upper arc and oblique structure are associated with the Mach stem and transverse wave, respectively. A kinked feature can be seen at their junction, ostensibly at the triple point. An incident shock located azimuthally upstream of the triple point (i.e., to the left of the wave in the projections) is expected from the standard structure, but it is not visible in the images. This suggests very weak or very delayed heat release behind the incident shock prior to the passage of the transverse wave. This interpretation is consistent with the results of Huang et al.~\cite{Huang2000}, who observed that the reaction zone behind the incident wave is spatially separated from the leading front, whereas heat release behind the Mach stem and transverse wave is coupled more closely to the shock structure.\par

\begin{figure}[ht]
    \centering
    \includegraphics[width=.5\textwidth]{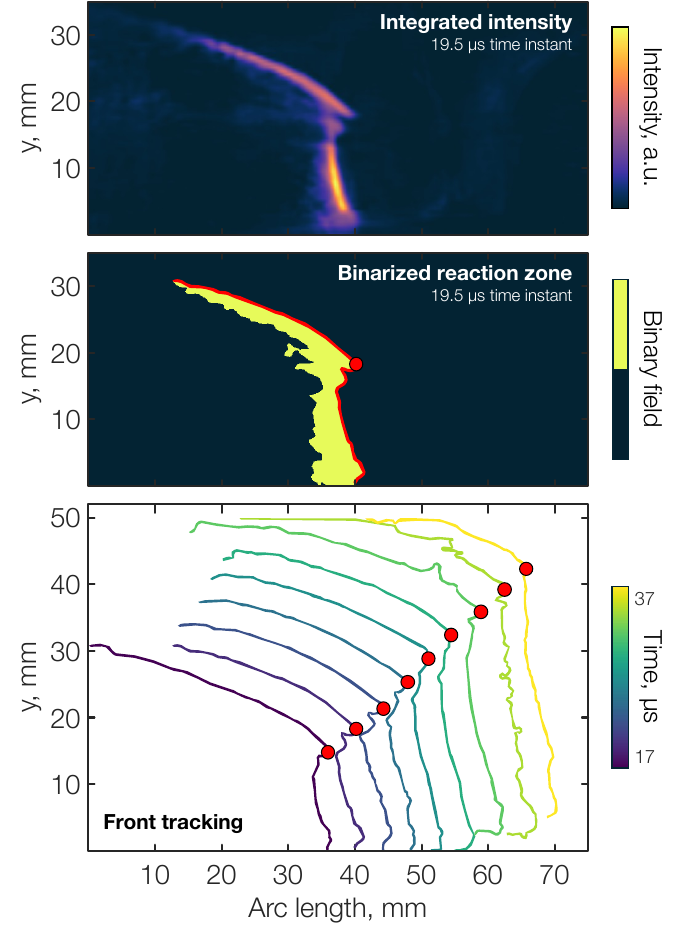}
    \caption{Wave parameter extraction: unwrapped intensity field (top), binarized structure with tracked front and kink location (middle), and time series of tracked wave fronts (bottom).}
    \label{fig: binarization}
\end{figure}

\begin{figure}[ht]
    \centering
    \includegraphics[width=.5\textwidth]{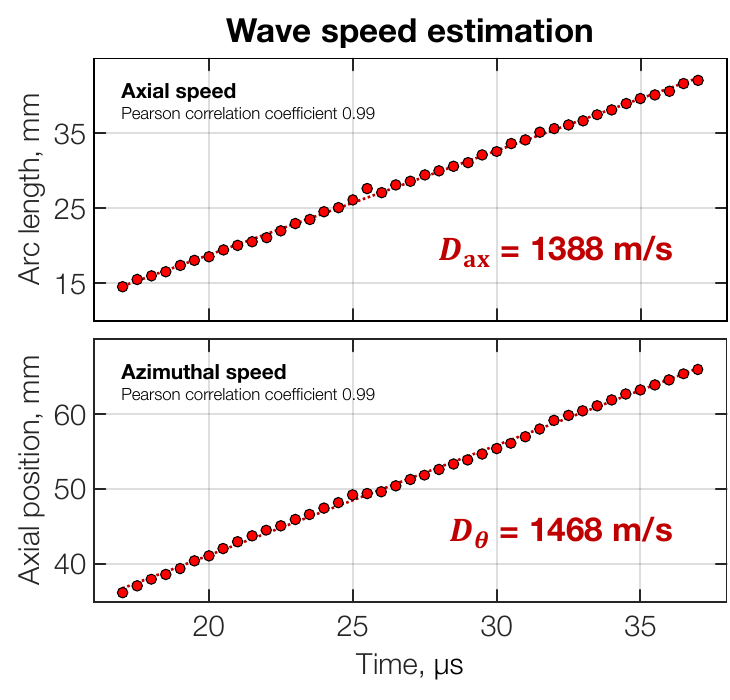}
    \caption{Wave-speed estimation for Case~A over from $t = 17$~$\upmu$s to 37~$\upmu$s: axial displacements over time (top) and azimuthal displacements (bottom).}
    \label{fig: wave speeds}
\end{figure}

Wave structure parameters for the persistent spinning wave can be extracted from the reconstructed time series. Because the intensity is concentrated at the wall, we focus on the outermost shell corresponding to $r = 31$~mm to 33~mm. Emission in this shell is radially averaged and ``unwrapped,'' as shown for the $t = 19.5~\upmu$s instant in the top panel of Fig.~\ref{fig: binarization}. To track the wave front, these images are normalized by their maximum values and binarized using a threshold of 0.1, resulting in profiles like those in the middle panel of Fig.~\ref{fig: binarization}. The binarized images are then used to extract the wave boundary and kink location (white ring), which serves as a proxy for the location of the triple point. The bottom panel of Fig.~\ref{fig: binarization} shows tracked wave fronts from $t = 17~\upmu$s to 37~$\upmu$s, plotted at every 2.5~$\upmu$s. These tracked fronts show a consistent global wave structure, but the detailed shape changes from frame to frame. The tracked boundaries are used to estimate the axial and azimuthal wave speeds, the track angle $\beta_1$, and the local flow angle $\beta_2$.\par

Figure~\ref{fig: wave speeds} plots the azimuthal and axial displacements of the kink location over time, loosely interpreted as the triple point trajectory. Linear fits to the tracked positions give axial and azimuthal speeds of $D_\mathrm{ax} = 1388$~m/s and $D_\theta = 1468$~m/s. This axial speed is quite close to that measured by the ion probes, i.e., 1364~m/s, corroborating the tomography-based tracking method. Moreover, both values are highly consistent over the observation window, as can be appreciated from the minimal scatter in Fig.~\ref{fig: wave speeds}. Indeed, the linear fits in this figure yield Pearson correlation coefficients of 0.99 for both $D_\mathrm{ax}$ and $D_\theta$. Using Eq.~\eqref{equ: spinning track angle}, the inferred velocity components give a track angle of $\beta_1 = 46.6$\textdegree\ and a pitch of $P = 2.97D$. Both of these values are within range of previously reported values for spinning detonations \cite{Huang2000, Tsuboi2007}.\par

\begin{figure}[ht]
    \centering
    \includegraphics[width=.5\textwidth]{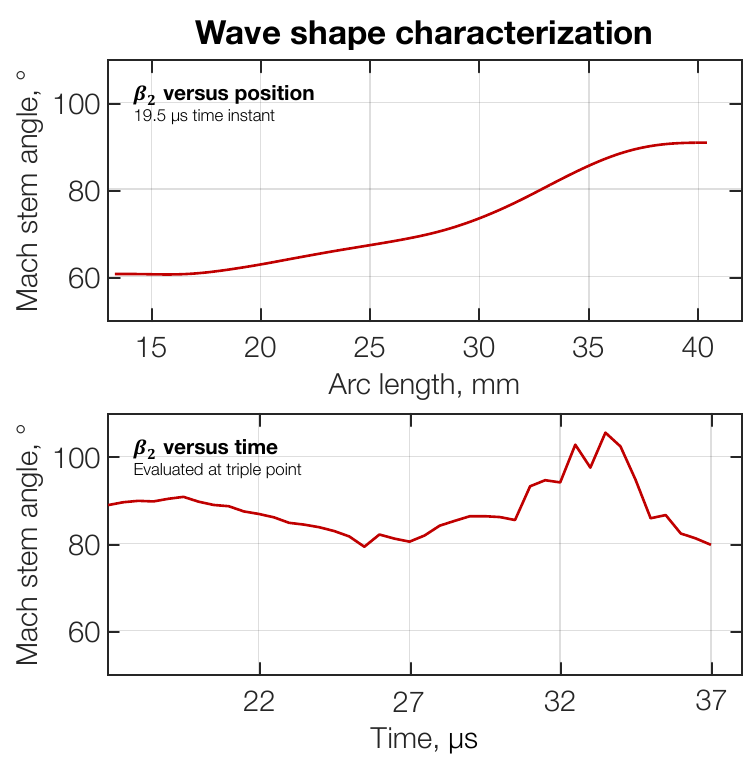}
    \caption{Mach stem angles for Case~A: local angle along the tracked front at $t = 19.5~\upmu$s (top) and distribution of $\beta_2$ near the triple point from $t = 17$~$\upmu$s to 37~$\upmu$s (bottom).}
    \label{fig: Mach stem angle}
\end{figure}

The angle $\beta_2$ is estimated from the leading front, which is presumed to be associated with the Mach stem. The front coordinates, expressed in terms of circumferential distance and axial position relative to the kink location, are fitted using a spline. The local slope of this fitted front is then used to estimate the angle between the incoming flow and the Mach stem in a shock-attached frame of reference. The top panel of Fig.~\ref{fig: Mach stem angle} shows the spatial distribution of $\beta_2$ along the front at $t = 19.5~\upmu$s. The angle approaches $90$\textdegree\ near the kink location, which is similar to the findings of Huang et al.~\cite{Huang2000}. The bottom panel of Fig.~\ref{fig: Mach stem angle} shows the estimated $\beta_2$ values near the triple point region over time, from $t = 17~\upmu$s to 37~$\upmu$s. Throughout most of this interval, $\beta_2$ is observed to fall between 80\textdegree\ and 90\textdegree, suggesting the presence of a strong, nearly normal Mach stem. Some frames yield values above 90\textdegree, however. These deviations are attributed to transient changes in the local wave shape, particularly when the front becomes rounded near the kink rather than forming a sharp bend. Nonetheless, these results show the potential for chemiluminescence tomography to provide quantitative assessments of detonation wave kinematics and morphology.\par

\subsection{Case~B: failed spinning detonation}
\label{sec: results: failed-spin}
\begin{figure}[ht]
    \centering
    \includegraphics[width=.5\textwidth]{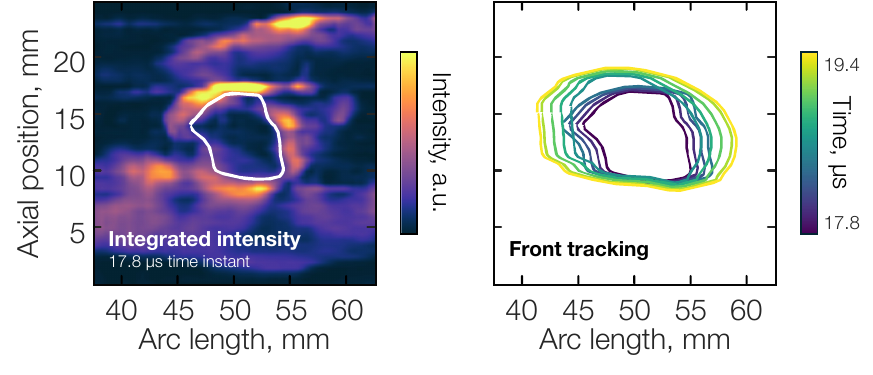}
    \caption{Wall-attached ring-like feature observed after failure in Case~B: feature boundary at $t = 17.8~\upmu$s (left) and tracked feature over time (right).}
    \label{fig: feature tracking}
\end{figure}

Case~B concerns a fuel-rich spinning detonation that fails within the test section. As introduced in Sec.~\ref{sec: campaign: cases}, the early projection images show a spinning structure, but the wave fragments between the frames recorded at 9.8~$\upmu$s and 19.8~$\upmu$s, as can be seen in the supplemental videos. Failure is characterized by the loss of a coherent spinning wave and a marked reduction in the chemiluminescence intensity. After failure, several intense ring-shaped features appear to emanate from the wall, superimposed on distributed emission throughout the tube. These structures are difficult to isolate in projection images because multiple emitting regions overlap along each line of sight. Figure~\ref{fig: feature tracking} shows one such feature. As in the persistent spinning case, the near-wall emission is obtained by averaging over a thin radial shell at the periphery of the domain. The reconstruction localizes this feature on the wall and separates it from the surrounding emission, enabling temporal tracking of the ring.\par

The feature boundary is manually initialized and evolved using an active-contour segmentation algorithm. Specifically, the inner boundary of the feature is drawn by hand, and MATLAB's \texttt{activecontour} function is used with the edge method to fit the ring structure. The right panel of Fig.~\ref{fig: feature tracking} shows the tracked feature boundary from $t = 17.8~\upmu$s to 19.4~$\upmu$s. Because the feature undergoes a substantial change in shape around 18.6~$\upmu$s, the segmentation is reinitialized at that time before tracking is continued. The average radial growth rate of the tracked feature over this interval is approximately $1720$~m/s, i.e., slightly in excess of the computed CJ speed. The ring expands predominantly along the azimuthal direction, where there is additional confinement due to the cylinder curvature. Characterizing these features from a 2D projection---wherein the orientation of the ring structure and direction of its expansion can be ambiguous---poses a significant challenge. However, the task is relatively straightforward via chemiluminescence tomography.\par

\subsection{Case~C: counter-propagating transverse fronts}
\label{sec: results: counter-propagating}
\begin{figure*}[t]
    \centering
    \includegraphics[width=.9\textwidth]{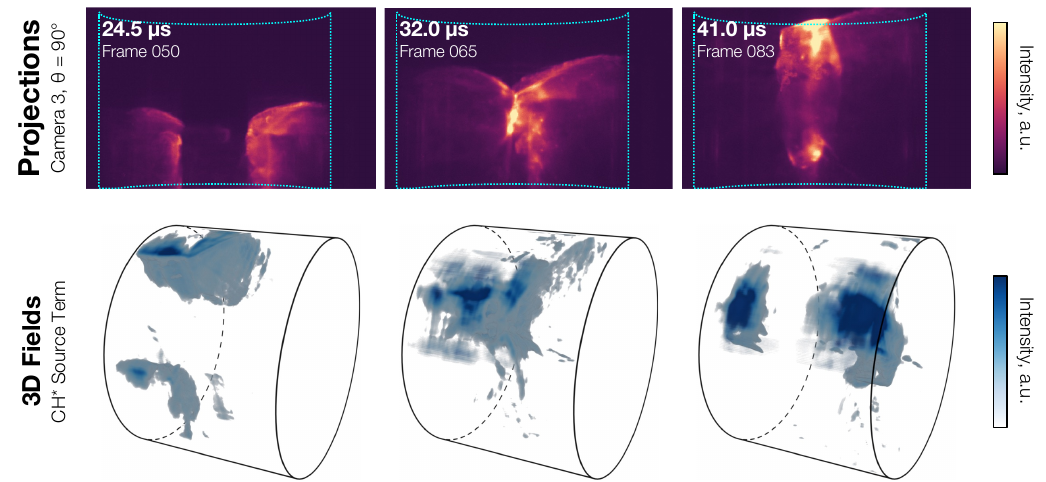}
    \caption{Projections (top) and reconstructions (bottom) of counter-propagating transverse fronts in a lean \ce{C2H4}--air mixture. Dashed teal lines superimposed on the projections indicate the boundary of the reconstruction volume.}
    \label{fig: reconstructions: case C}
\end{figure*}

The detonation wave in Case~C has two transverse fronts that propagate toward one another and collide within the test section. This case is useful for assessing the ability of chemiluminescence tomography to separate structures that overlap in the projections. Figure~\ref{fig: reconstructions: case C} shows reconstructed fields at $t = 19.5~\upmu$s, 32.0~$\upmu$s, and 41.0~$\upmu$s. The first frame captures the two fronts before the collision, and the second frame shows the interaction of these fronts. During the collision, the emission becomes much more intense. Afterwards, the region of intense emission fragments into two primary blobs that propagate axially, with a relatively fast-moving blob upstream of a relatively slow-moving one, as shown in the supplemental videos.\par

An approximate velocity is obtained for the stronger luminous front prior to the collision using the same near-wall tracking procedure detailed in Sec.~\ref{sec: results: failed-spin}. (The weaker front is less uniform and is not amenable to the same procedure.) The axial and azimuthal speeds of this front are estimated to be 1462~m/s and 1783~m/s, respectively. The axial component is below both the computed CJ speed of 1671~m/s and the experimentally measured wave speed of 1560~m/s. The lower axial speed inferred from tomography may indicate local deceleration of the wave prior to the collision. However, since the ion probes provide an average speed outside the visible test section, this discrepancy cannot be resolved from the present data. Conversely, the azimuthal speed is slightly above the CJ speed, which could be explained in part by intense conditions behind the incident wave \cite{Frederick2025} and a high effective activation energy \cite{Lu2021}, i.e., 16, as compared to 8.1 and 7.2 for cases~A and B.\par

\section{Conclusions}
\label{sec: conclusions}
This study demonstrates time-resolved chemiluminescence tomography of detonation waves in a round tube. Using five high-speed cameras, a refractive model of the imaging geometry, and a trilinear basis in cylindrical coordinates, we reconstructed 3D chemiluminescence fields over 128 frames for three representative detonations. Through assessment of these reconstructions, we show that chemiluminescence tomography can recover reacting structures that are difficult to isolate from individual line-of-sight integrated projection data.\par

For the persistent spinning detonation, the reconstructed emission field was consistent with the standard description of a single-head spinning wave. The leading arc and oblique structure were associated with the Mach stem and transverse wave, respectively, while the incident shock was not directly visible in the chemiluminescence data. This absence is consistent with weak or spatially delayed heat release behind the incident shock prior to the passage of the transverse wave, as suggested by prior studies. The reconstructions also enabled quantitative estimates of the wave kinematics. Feature tracking gave an axial speed in agreement with the value measured by an ion probe train; a track angle of $\beta_1 = 46.62$\textdegree\ and pitch of $P = 2.97D$ were deduced from the reconstructions, both being consistent with prior observations of spinning detonations. The local Mach stem angle $\beta_2$ was generally between 80\textdegree and 90\textdegree, supporting the interpretation of a strong, nearly normal Mach stem.\par

The two non-canonical cases further demonstrated the utility of the tomographic measurement. In the failed spinning case, the reconstruction localized ring-like features emanating from the wall that were superimposed with distributed emission within the test section, allowing their growth to be tracked. In the case with counter-propagating transverse fronts, the reconstruction separated post-collision structures that were convoluted with each other in the 2D projections.\par

The present reconstructions are limited by the small number of views and by the fact that chemiluminescence measures light emission associated with intermediate reactions and not the shock structure. As a result, weakly emitting regions, including the wake of the incident shock, may not be detectable. Future work should improve the fidelity of such reconstructions through additional viewing angles, where practical, as well as physics-based data assimilation that incorporates full or partial constraints from the governing equations. Such extensions would provide a path toward a more complete characterization of detonation structure, potentially allowing for the inference of uncertain model parameters.\par

\section*{Acknowledgments}
The authors thank Dr. Robert T. Fievisohn for his valuable technical input. This work was partially supported by the AFOSR\slash AFRL Center of Excellence in Assimilation of Flow Features in Compressible Reacting Flows under award no.~FA9550-25-1-0011, monitored by Dr. Chiping Li. Support for testing was provided by AFRL under prime contract no.~FA8650-22-F-2011 and by Innovative Scientific Solutions, Inc. under subcontract no.~SB20314.\par


\end{document}